# Transient Electroosmosis of a Maxwell fluid in a Rotating Microchannel

P Kaushik, Shubhadeep Mandal and Suman Chakraborty[*]

*Department of Mechanical Engineering,
Indian Institute of Technology Kharagpur, Kharagpur, 721302, India*

The transient electroosmotic flow of Maxwell fluid in a rotating microchannel is investigated both analytically and numerically. We bring out the complex dynamics of the flow during the transience due to the combination of rotation and rheological effects. We show the regimes of operation under which our analysis holds the most significance. We also shed some light on the volumetric flow rate characteristics as dictated by the underlying flow physics. Analytical solution compares well with the numerical solution. We believe that the results from the present study could potentially have far reaching applications in bio-fluidic microsystems where fluids such as blood, mucus and saliva may be involved.

[*] E-mail address for correspondence: suman@mech.iitkgp.ernet.in

## I. INTRODUCTION

The increasing importance of lab-on-a-CD based medical diagnostics has enhanced the interest in studying flows of bio-fluids in rotationally actuated microfluidic devices.[1–3] Flows of non-Newtonian liquids in such devices are complex to analyze, because of the inherent coupling between rotation-based effects such as the Coriolis force with the flow rheology.[4–7] Such a dynamic coupling tends to bring novel flow physics into the fray, which is of much interest. Along with the novel physics, augmented mixing also may be observed for certain types of rheological behaviour.[1,8–12] Augmented mixing is a desired effect in such micro-devices where the flow Reynolds number may be quite small. In order to have further augmentation in mixing, often an additional forcing, such as electrical actuation is considered along with rotational actuation (Coriolis effect). Such studies for augmented mixing has been of interest in recent times.[13–15]

Pure electroosmotic flow (without rotational effects) of rheologically complex fluids such as Maxwell fluids in itself becomes complex in theorizing and analyzing. Such pure electroosmotic flows of Maxwell fluids in parallel plate microchannels[12,16] and in rectangular microchannels[17] without rotational effects have been studied and well documented. When it comes to the flow of complex fluids under rotation, the effect of rheology on electroosmotic flow in rotating microchannels has been studied by few groups for power-law fluids,[18] third grade fluids[19] as well as viscoelastic fluids considering Oldroyd-B model.[20] The studies for power-law fluids[18] and viscoelastic fluids[20] have attempted numerical solutions, whereas the study with the solution for a third grade fluid[19] has been both analytical and numerical. The studies where rheological effects have been coupled with an electroosmotically driven flow in rotating microchannels have produced results of great interest, which includes augmented mixing.

Considering this, we attempt to analytically investigate the electroosmotic flow of Maxwell fluids in a rotating microfluidic channel and compare the same with numerical results from our in-house code. Electrical double layers (EDLs) are assumed to be thick but non-overlapping. Unique feature of the present investigation is its pure analytical nature, which, for the first time, brings out a closed form solution depicting the intricate coupling between electrokinetics, rotational hydrodynamics and viscoelastic nature of the flow medium.

## II. PROBLEM DESCRIPTION AND MATHEMATICAL FORMULATION OF THE PROBLEM

Unsteady flow of Maxwell fluid in a rotating microchannel is considered, as depicted in Fig. 1. The coordinate system is assumed to be rotating along with the channel, where the coordinate axes along the $x$, $y$ and $z$ represent the directions along the length, width and height of the microchannel respectively. We assume the channel to be rotating about the $z$ axis with a constant angular velocity $\mathbf{\Omega} = (0, 0, \Omega)$ as shown in Fig. 1. For this analysis, the length and width of the channel are assumed to be much larger than the height of the channel. A uniform electric field $\mathbf{E} = (E_x, 0, 0)$ is applied in the $x$-direction, as shown in Fig. 1. The fluid is assumed to be at rest initially and free of any residual stresses. The height of the channel is considered to be $2H$. The rotational effect along $z$-direction with electroosmotic effect along $x$-direction is considered in further study.

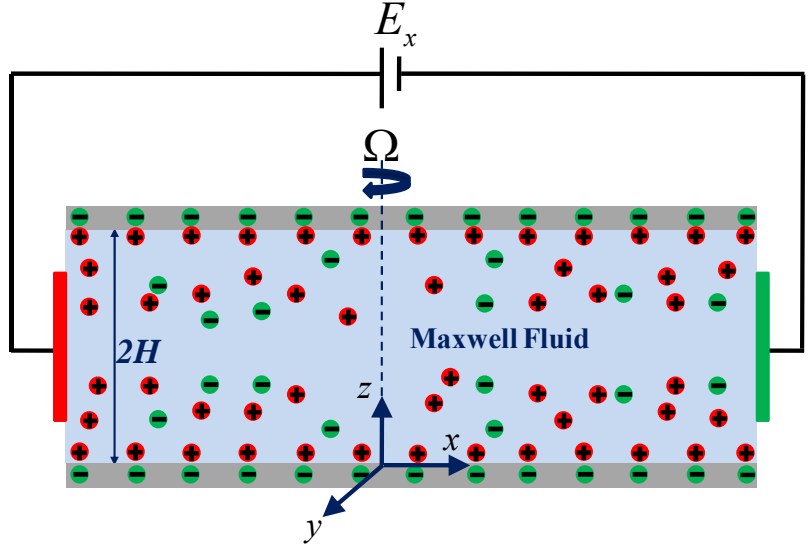

FIG. 1 Schematic representation of the physical setup under consideration.

## A. The velocity distribution in the flow field

The condition of incompressibility and the Cauchy equation of fluid motion, governing the motion of a fluid in a rotating reference frame is given as [21–23]:

$$\nabla \cdot \mathbf{u} = 0, \tag{1}$$

$$\rho \left\{ \frac{\partial \mathbf{u}}{\partial t} + (\mathbf{u} \cdot \nabla)\mathbf{u} + 2(\mathbf{\Omega} \times \mathbf{u}) + \mathbf{\Omega} \times (\mathbf{\Omega} \times \mathbf{r}) \right\} = -\nabla p + \nabla \cdot \mathbf{T} + \rho_e \mathbf{E}. \tag{2}$$

The term $\mathbf{T}$ represents the deviatoric stress tensor, $\rho_e \mathbf{E}$ represents electroosmotic body force per unit volume ($\rho_e$ is the net electric charge density and $\mathbf{E}$ is the externally applied axial electric field), $\mathbf{u} = (u, v, w)$ represents the velocity field, $\mathbf{r}$ represents the position vector measured from rotational axis and $\mathbf{\Omega}$ represents the angular velocity vector. For a upper convected Maxwell fluid, the deviatoric stress tensor is given as:[24]

$$\mathbf{T} + \lambda_1 \overset{\nabla}{\mathbf{T}} = 2\eta \mathbf{D}, \tag{3}$$

where $\overset{\nabla}{\mathbf{T}}$ is the upper convected derivative of $\mathbf{T}$ and $\eta$ is the dynamic viscosity of the fluid.

$\overset{\triangledown}{\mathbf{T}}$ has the following form:[25]

$$\overset{\triangledown}{\mathbf{T}} = \frac{\partial \mathbf{T}}{\partial t} + (\mathbf{u} \cdot \nabla) \cdot \mathbf{T} - (\nabla \mathbf{u}) \cdot \mathbf{T} - \mathbf{T} \cdot (\nabla \mathbf{u})^T. \tag{4}$$

**D** is the rate-of-strain tensor defined as

$$\mathbf{D} = \frac{1}{2}\{\nabla \mathbf{u} + (\nabla \mathbf{u})^T\}. \tag{5}$$

The components of **T** appearing in Eq. (4) can be represented in matrix form as follows:

$$\mathbf{T} = \begin{bmatrix} \tau_{xx} & \tau_{xy} & \tau_{xz} \\ \tau_{yx} & \tau_{yy} & \tau_{yz} \\ \tau_{zx} & \tau_{zy} & \tau_{zz} \end{bmatrix}. \tag{6}$$

We would like to mention here that we use the modified pressure $P = p - \left(\rho |\mathbf{\Omega} \times \mathbf{r}|^2 / 2\right)$ in order to simplify Eq. (2), to include the centrifugal term in the pressure term.[26] Simplified form of Eq. (2), with modified pressure term may be written as:

$$\rho\left\{\frac{\partial \mathbf{u}}{\partial t} + (\mathbf{u} \cdot \nabla)\mathbf{u} + 2(\mathbf{\Omega} \times \mathbf{u})\right\} = -\nabla P + \nabla \cdot \mathbf{T} + \rho_e \mathbf{E}. \tag{7}$$

**B. Simplification of charge distribution in the channel**

Determination of the electrical body force term $\rho_e \mathbf{E}$ present in Eq. (7) requires the charge distribution inside the channel. Charge distribution $(\rho_e)$ and electric potential distribution $(\psi)$ within the EDL are governed by Poisson equation of the form [27,28]:

$$\varepsilon \frac{d^2\psi}{dz^2} = -\rho_e. \tag{8}$$

Here, the net ionic charge density is represented by $\rho_e$. For a $\varsigma:\varsigma$ symmetric electrolyte, $\rho_e$ may be written as $\rho_e = e\varsigma(n^+ - n^-)$, where $\varsigma$ is the valence of the ions. The co-ion and counter-ion number densities in the EDL is represented by $n^\pm$. This is obtained from the Boltzmann distribution, which may be written as $n^\pm = n_\infty \exp[\mp e\varsigma\psi / k_B T]$, with $n_\infty$ representing the ionic concentration of the bulk, $e$ is the charge of a proton. $T$ and $k_B$ are the absolute temperature and the Boltzmann constant respectively. In the present study we assume that the wall potential $(\psi_w)$ also known as the zeta potential to be small $(\leq 25 \text{ mV})$, using which, the Debye-Hückel linearization[27,29] maybe considered. Eq. (8) may therefore be simplified as:

$$\frac{d^2\psi}{dz^2} = \kappa^2 \psi, \tag{9}$$

where $\kappa$ represents the inverse of the EDL thickness with $\kappa^2 = (2n_\infty e^2 \varsigma^2)/(\varepsilon k_B T)$. The solution of Eq. (9) is obtained using the following boundary conditions:

$$\left. \begin{array}{l} \text{at } z = 0, \ \psi = \psi_w, \\ \text{at } z = 2H, \ \psi = \psi_w, \end{array} \right\} \tag{10}$$

where $\psi_w$ is the specified wall potential. The solution for the potential distribution within the EDL as described in Eq. (9) with the boundary conditions shown in Eq. (10) can be obtained as:

$$\psi(z) = \psi_w \left[ \frac{\cosh\{\kappa(z-H)\}}{\cosh(\kappa H)} \right]. \tag{11}$$

The charge density can be obtained as

$$\rho_e = -\varepsilon \psi_w \kappa^2 \left[ \frac{\cosh\{\kappa(z-H)\}}{\cosh(\kappa H)} \right]. \tag{12}$$

Using Eq. (12), the electrical body force term in the governing momentum equation [Eq. (7)] is simplified and written as,

$$\rho\left\{\frac{\partial \mathbf{u}}{\partial t} + (\mathbf{u} \cdot \nabla)\mathbf{u} + 2(\mathbf{\Omega} \times \mathbf{u})\right\} = -\nabla P + \nabla \cdot \mathbf{T} - \varepsilon \psi_w \kappa^2 \left[ \frac{\cosh\{\kappa(z-H)\}}{\cosh(\kappa H)} \right] \mathbf{E}. \tag{13}$$

**C. Simplification of the momentum equation**

It has been assumed that for the channel, the length $\gg$ width $\gg$ height of the channel. This automatically means that $\frac{\partial}{\partial x}(\ ) \ll \frac{\partial}{\partial y}(\ ) \ll \frac{\partial}{\partial z}(\ )$ and the terms containing $\frac{\partial}{\partial x}(\ )$, $\frac{\partial^2}{\partial x^2}(\ )$, $\frac{\partial}{\partial y}(\ )$ and $\frac{\partial^2}{\partial y^2}(\ )$ are dropped from Eq. (13) as they can be ignored in comparison to the terms containing $\frac{\partial}{\partial z}(\ )$ and $\frac{\partial^2}{\partial z^2}(\ )$. We have also assumed that the fluid is at rest initially and is unstressed. These considerations (including no-slip and no-penetration at the walls) reduce the value of $w$ throughout the domain to be zero and also $\tau_{zz}$ remains zero throughout. The desired solution for $u$ and $v$ for a very long channel makes the velocities into a function of $z$ and $t$ only and makes modified pressure gradient terms $\frac{\partial P}{\partial x}$ and $\frac{\partial P}{\partial y}$ ignorable compared to other terms in the equation [14,19,20,30]. Therefore, we can set $w = 0$, $\tau_{zz} = 0$ and $\frac{\partial w}{\partial z} = 0$, while dropping the modified pressure gradient terms in the governing equations, and they may be reduced to,

$$\rho\left(\frac{\partial u}{\partial t}-2\Omega v\right)=\frac{\partial \tau_{xz}}{\partial z}+\rho_e E_x, \tag{14}$$

$$\rho\left(\frac{\partial v}{\partial t}+2\Omega u\right)=\frac{\partial \tau_{yz}}{\partial z}, \tag{15}$$

$$\frac{\partial P}{\partial z}=0, \tag{16}$$

$$\tau_{xx}+\lambda_1\left(\frac{\partial \tau_{xx}}{\partial t}-2\tau_{zx}\frac{\partial u}{\partial z}\right)=0, \tag{17}$$

$$\tau_{xy}+\lambda_1\left(\frac{\partial \tau_{xy}}{\partial t}-\tau_{yz}\frac{\partial u}{\partial z}-\tau_{xz}\frac{\partial v}{\partial z}\right)=0, \tag{18}$$

$$\tau_{xz}+\lambda_1\left(\frac{\partial \tau_{xz}}{\partial t}\right)=\eta\frac{\partial u}{\partial z}, \tag{19}$$

$$\tau_{yy}+\lambda_1\left(\frac{\partial \tau_{yy}}{\partial t}-2\tau_{yz}\frac{\partial v}{\partial z}\right)=0, \tag{20}$$

$$\tau_{yz}+\lambda_1\left(\frac{\partial T_{23}\tau_{yz}}{\partial t}\right)=\eta\frac{\partial v}{\partial z}. \tag{21}$$

We observe from Eqs. (14)-(21) that Eqs. (17), (18) and (20) do not have any effect on the flow described by Eqs. (14)-(16). Also, equation (16), describing the $z$-momentum equation may be dropped by setting $P(t,z)=0$ throughout.

### D. Non-dimensionalization of the transport equations

We seek the dimensionless set of equations and boundary conditions in order to further simplify the governing equations. Here, we use channel half-height $H$ as reference length scale, the Helmholtz-Smoluchowski (HS) velocity, $u_{HS}=-\varepsilon E_x \psi_w/\eta$ as the reference velocity scale and $\rho H^2/\eta$ as the reference time scale. Non-dimensionalization of the transport equations is done by invoking the following dimensionless variables: $z^*=\frac{z}{H}$, $\psi^*=\frac{\psi}{\psi_w}$, $u^*=\frac{u}{u_{HS}}$, $t^*=\frac{t\eta}{\rho H^2}$, and $\mathbf{T}^*=\frac{\mathbf{T}H}{\eta u_{HS}}$. This non-dimensional scheme leads to description of the physical system in terms of the following dimensionless numbers: $\mathrm{El}=\frac{\lambda_1 \eta}{\rho H^2}$, $\kappa^*=\kappa H$ and $\mathrm{Re}_\Omega=\frac{2\rho\Omega H^2}{\eta}$. Here, $\kappa^*$ represents the inverse of dimensionless

EDL thickness, $Re_\Omega$ represents the rotational Reynolds number[20] and El represents the elasticity number.[31] We, therefore obtain the dimensionless governing differential equations as:

$$\frac{\partial u^*}{\partial t^*} - Re_\Omega v^* = \frac{\partial \tau_{xz}^*}{\partial z^*} + \kappa^{*2}\psi^*, \quad (22)$$

$$\frac{\partial v^*}{\partial t^*} + Re_\Omega u^* = \frac{\partial \tau_{yz}^*}{\partial z^*}, \quad (23)$$

$$\tau_{xz}^* + El\frac{\partial \tau_{xz}^*}{\partial t^*} = \frac{\partial u^*}{\partial z^*}, \quad (24)$$

$$\tau_{yz}^* + El\frac{\partial \tau_{yz}^*}{\partial t^*} = -\frac{\partial v^*}{\partial z^*}. \quad (25)$$

Eqs. (22) and (23) may be further arranged using Eqs. (24) and (25), and may be written as:

$$El\frac{\partial^2 u^*}{\partial t^{*2}} + \frac{\partial u^*}{\partial t^*} - El\,Re_\Omega \frac{\partial v^*}{\partial t^*} = Re_\Omega v^* + \frac{\partial^2 u^*}{\partial z^{*2}} + \kappa^{*2}\psi^*, \quad (26)$$

$$El\frac{\partial^2 v^*}{\partial t^{*2}} + \frac{\partial v^*}{\partial t^*} + El\,Re_\Omega \frac{\partial u^*}{\partial t^*} = -Re_\Omega u^* + \frac{\partial^2 v^*}{\partial z^{*2}}. \quad (27)$$

The initial and boundary conditions used for finding the solution of Eqs. (26)-(27) in dimensionless form may be written as:

$$\left.\begin{array}{l}\text{Initial conditions: } u^*\big|_{t=0}=0,\quad v^*\big|_{t=0}=0,\quad \frac{\partial u^*}{\partial t^*}\bigg|_{t^*=0}=0,\quad \frac{\partial v^*}{\partial t^*}\bigg|_{t^*=0}=0,\\ \text{Boundary conditions: } u^*\big|_{z^*=0}=0,\quad v^*\big|_{z^*=0}=0,\quad u^*\big|_{z^*=2}=0,\quad v^*\big|_{z^*=2}=0.\end{array}\right\} \quad (28)$$

It is important to mention here that for a Newtonian fluid $(i.e.\ El=0)$, unsteady analytical solution of Eqs. (26)-(27) was found by Gheshlaghi et al.[15] and steady state solution was found by Chang and Wang.[32] Here, we progress to find analytical solution for Eqs. (26)-(27) and also for a special case of Newtonian fluid we validate our solution with Gheshlaghi et al.[15] We would like to mention here that, hereon, we drop the superscript '*' in the equations and boundary conditions for ease of presentation and symbols without superscript '*' represent the dimensionless quantities.

## III. SOLUTION

### A. Analytical Solution

Towards progressing with analytical treatment, it is convenient to introduce a complex function $\chi(z,t) = u(z,t) + iv(z,t)$, where $i = \sqrt{-1}$. Using the definition of $\chi(z,t)$, one can combine Eqs. (26) and (27) to obtain the following complex PDE

$$\text{El}\frac{\partial^2 \chi}{\partial t^2} + \frac{\partial \chi}{\partial t} + i\text{ElRe}_\Omega \frac{\partial \chi}{\partial z} + i\text{Re}_\Omega \chi - \frac{\partial^2 \chi}{\partial z^2} = \kappa^2 \psi. \tag{29}$$

The initial and boundary conditions in terms of the complex variable $\chi(z,t)$ can be obtained as

$$\left.\begin{array}{l}\text{Initial conditions: } \chi(z,t)\big|_{t=0} = 0, \quad \dfrac{\partial \chi(z,t)}{\partial t}\bigg|_{t=0} = 0, \\ \text{Boundary conditions: } \chi(z,t)\big|_{z=0} = 0, \quad \chi(z,t)\big|_{z=2} = 0.\end{array}\right\} \tag{30}$$

A closer look into Eqs. (29) and (30) reveals that the governing differential equation and associated subsidiary conditions (initial and boundary) are linear. This linearity allows us to express $\chi(z,t)$ in terms of superposition of two functions $\chi_1(z)$ and $\chi_2(z,t)$ in the following form

$$\chi(z,t) = \chi_1(z) + \chi_2(z,t), \tag{31}$$

where $\chi_1(z)$ and $\chi_2(z,t)$ represent the steady state and transient components of $\chi(z,t)$ respectively. Substituting Eq. (31) into Eq. (29), we obtain the governing differential equation for $\chi_1(z)$ as

$$\frac{d^2 \chi_1}{dz^2} - i\text{Re}_\Omega \chi_1 = -\kappa^2 \psi, \tag{32}$$

subject to the following boundary conditions:

$$\left.\begin{array}{l}\chi_1(z)\big|_{z=0} = 0, \\ \chi_1(z)\big|_{z=2} = 0.\end{array}\right\} \tag{33}$$

Similarly, we obtain the governing differential equation for $\chi_2(z,t)$ as

$$\text{El}\frac{\partial^2 \chi_2}{\partial t^2} + \frac{\partial \chi_2}{\partial t} + i\text{ElRe}_\Omega \frac{\partial \chi_2}{\partial z} + i\text{Re}_\Omega \chi_2 - \frac{\partial^2 \chi_2}{\partial z^2} = 0, \tag{34}$$

subject to the following initial and boundary conditions:

$$\left.\begin{array}{l}\text{Initial conditions: } \chi_2(z,t)\big|_{t=0} = -\chi_1(z), \quad \dfrac{\partial \chi_2(z,t)}{\partial t}\bigg|_{t=0} = 0, \\ \text{Boundary conditions: } \chi_2(z,t)\big|_{z=0} = 0, \quad \chi_2(z,t)\big|_{z=2} = 0.\end{array}\right\} \tag{35}$$

The general solution of $\chi_1(z)$ can be obtained by combining the complementary functions and particular integral as

$$\chi_1(z) = C_1 \cosh\left(\sqrt{i\mathrm{Re}_\Omega}\,z\right) + C_2 \sinh\left(\sqrt{i\mathrm{Re}_\Omega}\,z\right) + \frac{\kappa^2 \cosh\{\kappa(z-1)\}}{\cosh(\kappa)\left(i\mathrm{Re}_\Omega - \kappa^2\right)}, \tag{36}$$

where $C_1$ and $C_2$ are obtained by using the boundary conditions given in Eq. (33) as

$$C_1 = \frac{\kappa^2}{\kappa^2 - i\mathrm{Re}_\Omega},\ C_2 = -C_1 \coth\left(2\sqrt{i\mathrm{Re}_\Omega}\right) - \frac{\kappa^2}{\sinh\left(2\sqrt{i\mathrm{Re}_\Omega}\right)\left(i\mathrm{Re}_\Omega - \kappa^2\right)}. \tag{37}$$

Towards solving Eq. (34), we use the separation of variables method to represent $\chi_2(z,t)$ as

$$\chi_2(z,t) = Z(z)T(t), \tag{38}$$

where $Z(z)$ is a sole function of $z$ and $T(t)$ is a sole function of $t$. Substituting Eq. (38) into Eq. (34) and rearranging we obtain

$$\frac{1}{T}\left[\mathrm{El}\frac{d^2T}{dt^2} + \frac{dT}{dt} + i\mathrm{ElRe}_\Omega \frac{dT}{dt}\right] + i\mathrm{Re}_\Omega = \frac{1}{Z}\frac{d^2Z}{dz^2} = -\alpha^2, \tag{39}$$

with $\alpha^2$ being the separation constant. So, $Z(z)$ is governed by the following differential equation

$$\frac{d^2Z}{dz^2} + \alpha^2 Z = 0, \tag{40}$$

subject to the following boundary conditions:

$$\left.\begin{array}{l} Z(z)\big|_{z=0} = 0, \\ Z(z)\big|_{z=2} = 0. \end{array}\right\} \tag{41}$$

Eqs. (40) and (41) represent an eigenvalue problem, solution of which can be obtained in terms of the following eigen function

$$Z_n = a_n \sin(\alpha_n z), \tag{42}$$

where $a_n$ is an arbitrary constant and $\alpha_n$ is the eigenvalue of the form

$$\alpha_n = \frac{n\pi}{2} \quad \text{for } n = 1,2,3,\ldots \tag{43}$$

Now for each value of $n$, the governing differential equation for $T(t)$ can be written from Eq. (39) as

$$\frac{d^2T_n}{dt^2} + \left(\frac{1}{\mathrm{El}} + i\mathrm{Re}_\Omega\right)\frac{dT_n}{dt} + \frac{\left(\alpha^2 + i\mathrm{Re}_\Omega\right)}{\mathrm{El}}T_n = 0. \tag{44}$$

The general solution for $T_n(t)$ can be obtained in the following form

$$T_n(t) = b_n \exp(m_1 t) + c_n \exp(m_2 t), \tag{45}$$

where $b_n$ and $c_n$ are arbitrary constants. $m_1$ and $m_2$ in Eq. (45) are obtained as

$$m_1 = -\frac{1}{2}\left(\frac{1}{\text{El}} + i\text{Re}_\Omega\right) + \frac{1}{2}\sqrt{\left(\frac{1}{\text{El}} + i\text{Re}_\Omega\right)^2 - \frac{4}{\text{El}}\left(\alpha_n^2 + i\text{Re}_\Omega\right)}, \\ m_2 = -\frac{1}{2}\left(\frac{1}{\text{El}} + i\text{Re}_\Omega\right) - \frac{1}{2}\sqrt{\left(\frac{1}{\text{El}} + i\text{Re}_\Omega\right)^2 - \frac{4}{\text{El}}\left(\alpha_n^2 + i\text{Re}_\Omega\right)}. \tag{46}$$

So, the general solution for $\chi_2(z,t)$ can be obtained by combining Eqs. (42) and (45) as

$$\chi_2(z,t) = \sum_{n=1}^{\infty} T_n Z_n = \sum_{n=1}^{\infty} \{E_n \exp(m_1 t) + F_n \exp(m_2 t)\} \sin(\alpha_n z), \tag{47}$$

where $E_n = a_n b_n$ and $F_n = a_n c_n$ are the arbitrary constants. Substituting expressions of $\chi_1(z)$ and $\chi_2(z,t)$ in the two initial conditions (given in Eq. (35)), we obtain $E_n$ and $F_n$ as

$$E_n = -\left(\frac{m_2}{m_2 - m_1}\right)\int_0^2 \chi_1(z)\sin(\alpha_n z)dz, \\ F_n = -\frac{m_1}{m_2}E_n. \tag{48}$$

Combining Eqs. (36) and Eq. (47), we obtain the final solution for $\chi(z,t)$ as

$$\chi(z,t) = \begin{bmatrix} C_1 \cosh\left(\sqrt{i\text{Re}_\Omega}\,z\right) + C_2 \sinh\left(\sqrt{i\text{Re}_\Omega}\,z\right) + \dfrac{\kappa^2 \cosh\{\kappa(z-1)\}}{\cosh(\kappa)(i\text{Re}_\Omega - \kappa^2)} \\ + \sum_{n=1}^{\infty} \{E_n \exp(m_1 t) + F_n \exp(m_2 t)\} \sin(\alpha_n z) \end{bmatrix}. \tag{49}$$

Velocity components can be obtained as $u = \text{Re}(\chi)$ and $v = \text{Im}(\chi)$. To characterize the volume transport through the channel, we obtain the axial flow rate $Q_x(t)$, transverse flow rate $Q_y(t)$ and the angle of flow $\theta_f(t)$ in the following form

$$Q_x(t) = \int_0^2 u\,dz, \quad Q_y(t) = \int_0^2 v\,dz, \quad \theta_f(t) = \tan^{-1}\left(\frac{Q_y}{Q_x}\right). \tag{50}$$

**B. Numerical Solution**

We also obtain the numerical solution of Eqs. (26)-(27) along with boundary conditions delineated in Eqs. (28). We use an in-house finite difference code to find the numerical solution. The equations are discretized by second order central differencing in space and second order forward in time. The discretized equations are solved using SOR method.[33] We perform grid independence studies on our code and found the final mesh size to be of $10^4$ nodes. We also perform the time-step independence and chose our final time step as $\Delta t = 1\times 10^{-4}$. Numerical solution is hence used to affirm the correctness of the analytical solution and compare the same.

## IV. RESULTS AND DISCUSSION

We start the discussion with Figs. 2(a)-(b), which depict the evolution of the two velocity components in the channel with time. We show the matching between our analytical and numerical solutions as a benchmark. It is observed that there is good agreement between the two. The parameter values chosen for plotting Figs. 2(a)-(b) are $\kappa = 20$, $\mathrm{Re}_\Omega = 1$ and $\mathrm{El} = 1$. It is important to mention in this context that these values fall within the standard permissible limit.[32,34,35] It is seen from the Figs. 2(a)-(b) that the electroosmotic forcing which essentially acts in the zone close to the walls, drives the fluid close to the walls at small times $(t = 0.1)$. Maxwell fluids have a finite relaxation time, which makes the propagation of the fluid motion slower into the bulk of the fluid. It is also observed that the diffusivity is weaker at smaller times for a Maxwell fluid, thereby shapes with sharp discontinuities are quite common at smaller times (See Fig. 2(a) at $t = 1$). It can also be observed that the magnitude of the two velocities increases above the steady state value before decreasing and reaching steady state.

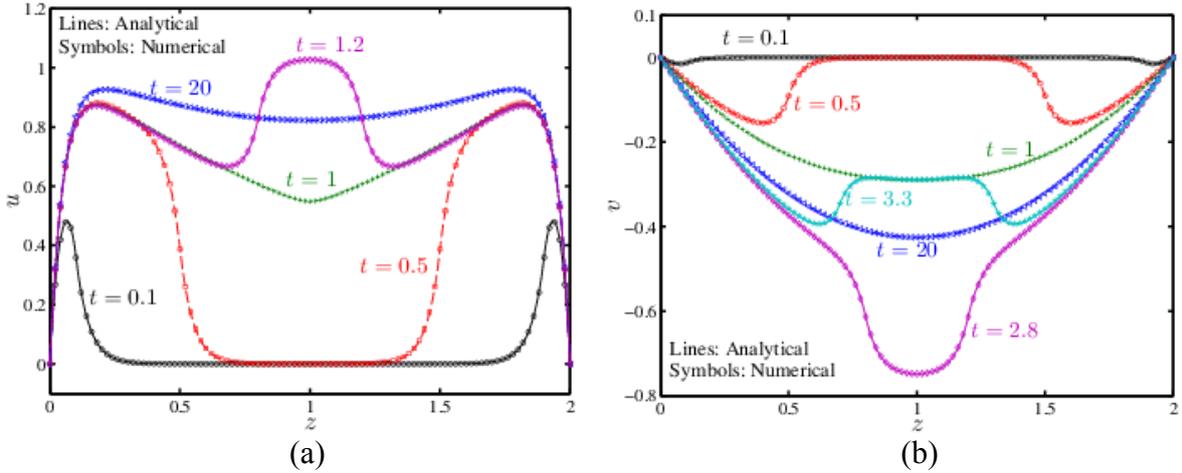

(a)          (b)

FIG. 2. Comparison of analytical and numerical solutions of (a) x velocity profile and (b) y velocity profile at different time for $\kappa = 20$, $\mathrm{Re}_\Omega = 1$ and $\mathrm{El} = 1$. Lines represent the analytically obtained solution, while symbols are used to represent the numerical solution.

In Fig. 2(a), we see that the $u$ velocity magnitude at the center of the channel is higher than steady state $(t = 20)$ value at $t = 1.2$. Similarly, the $v$ velocity magnitude is higher than the steady state value at $t = 2.8$ and then goes to a value lower than the steady state value at $t = 3.3$. This means that oscillatory nature of the flow becomes prominent for Maxwell fluids. These oscillations are due to the finite relaxation time scale of the fluid, which makes the fluid accept changes of the surrounding slower than for the case with zero relaxation time (Newtonian fluid). Importantly we observe that the steady state profile for both $u$ and $v$ velocities $(t=20)$ match the solution obtained by Chang and Wang[32] for a Newtonian fluid. This means that the difference between a Maxwell fluid and Newtonian fluid ceases to exist at steady state for the present case with the appropriate assumptions.

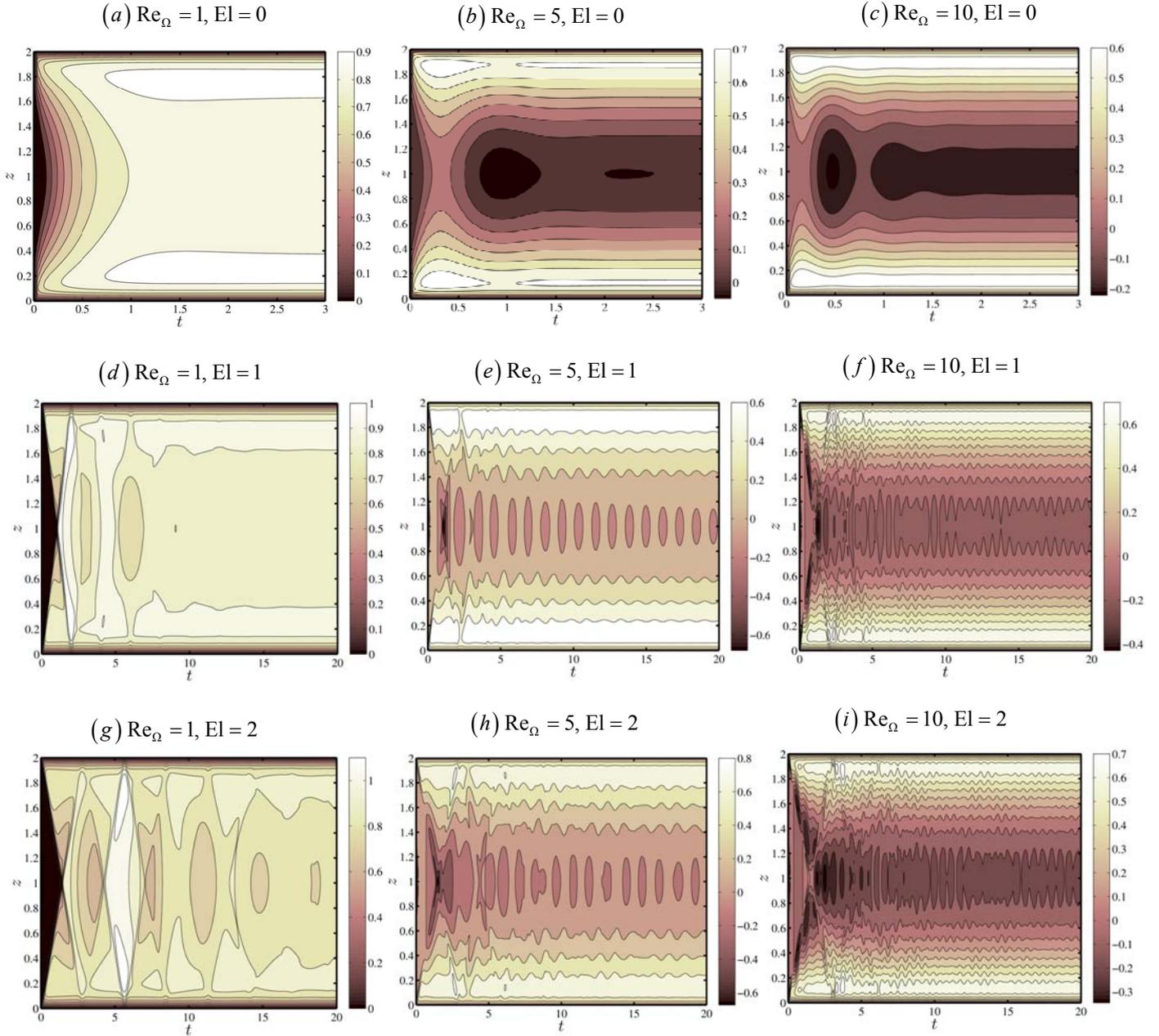

FIG. 3. Velocity evolution $u(z,t)$ for $\text{Re}_\Omega = 1, 5$ and $10$, $\text{El} = 0, 1$ and $2$ and $\kappa = 20$.

We further elaborate the effect of rotation and rheology during transience on the $u$ velocity in Figs. 3(a)-(i). The dimensionless inverse of EDL thickness chosen for this study $\kappa = 20$. We depict in Figs. 3(a)-(c) the cases with $\text{El} = 0$ and finite $\text{Re}_\Omega$. The cases with $\text{El} = 0$ corresponds to a Newtonian fluid. It is observed that our results show good match with results obtained for transience of a Newtonian fluid as studied by Gheshlaghi et al.[15] for both $u$ and $v$ (see Figs. 4(a)-(c)) velocity components. It can be seen from Figs. 3(a)-(c), that as $Re_\Omega$ increases, the transience time increases as well as the oscillation in the $u$ velocity magnitude at various location is larger. This is true for cases with higher El as well as clearly seen in Figs. 3(d)-(i). We observe that the oscillation of the $u$ velocity magnitude increases with increase in El. It is also observed that the $u$ velocity magnitude is negative for higher values of $Re_\Omega$, i.e. for $Re_\Omega = 10$ for $El = 0$, we observe negative values of $u$ velocity. When the value of El is increased to 1, we observe negative values for $u$ velocity at $Re_\Omega = 5$ itself. The oscillatory nature of Maxwell fluid coupled with oscillation caused by rotation allows for negative $u$ velocity to exist at certain locations away from the walls during transience (See Figs 3(e), (f), (h) and (i) between $t = 0$ and $t = 5$ in the zone close to $z = 1$). One may also observe that irrespective of the El, the oscillations near the walls are lesser than closer to the center of the channel. However, for the cases with higher values of El, oscillations are observed close to the wall at short times. This may be attributed to the existence of the wall, which causes a stabilizing effect on the flow close to it. However, in the case of the Maxwell fluid model, at shorter times the diffusive nature of the fluid is weaker, and as time progresses, the diffusive nature becomes stronger and the oscillations get damped.

In Figs. 4(a)-(i), we depict the evolution of $v$ velocity with time for various rotational Reynolds numbers and elasticity numbers. We observe from the Figs. 4(a)-(i) that the time taken to reach steady state increases with increase in El as well as $Re_\Omega$. We have earlier seen in Fig. 2(b) that the magnitude of $v$ velocity is essentially negative, but in cases with higher El and $Re_\Omega$, we observe that during transience the $v$ velocity magnitude becomes positive at some locations away from the walls. For example, in the case with $Re_\Omega = 10$ and $El = 2$, depicted in Fig. 4(i), it can be seen that $v$ velocity magnitude becomes positive at center of the channel between $t = 0$ and $t = 5$. In the case of $v$ velocity also, just like $u$ velocity, we observe more oscillatory nature of the velocity magnitude away from the walls. Close to the walls, these oscillations are damped quickly. We also observe that the oscillations are higher in amplitude as well as frequency for larger values of El and $Re_\Omega$. It may be also observed that with increase in El, the fluid becomes more elastic in nature and tends to have larger velocity magnitude during transience, and eventually settling down to steady state. For the case depicted in Fig. 4(g), we observe that the magnitude of velocity close to the center of the channel between $t = 0$ and $t = 5$ becomes larger than between $t = 5$ and $t = 10$. Eventually steady state is reached with velocity magnitudes lower than at some times during transience.

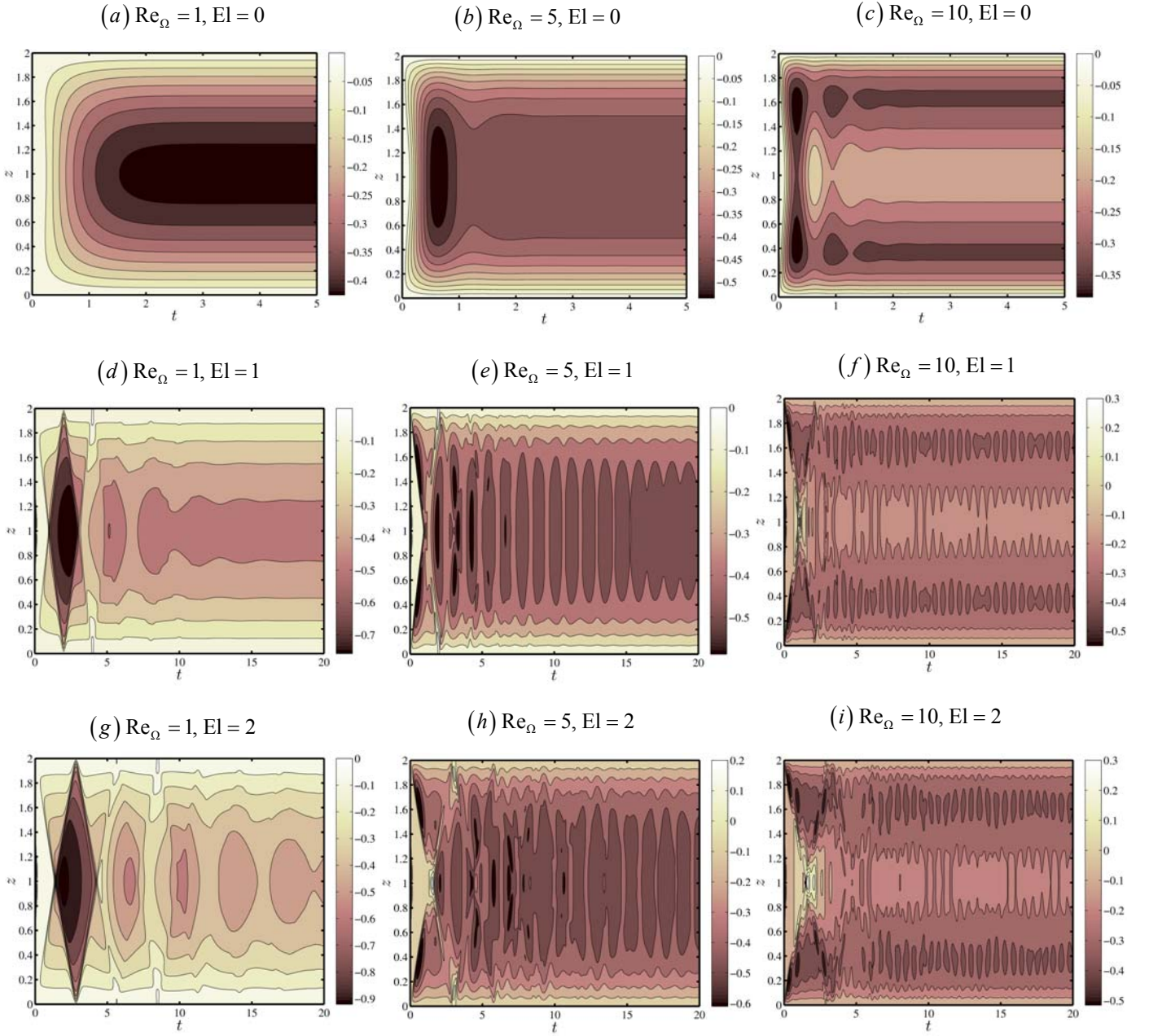

FIG. 4. Velocity evolution $v(z,t)$ for $\text{Re}_\Omega = 1, 5$ and $10$, $\text{El} = 0, 1$ and $2$ and $\kappa = 20$.

It is important to mention here that the combined effect of decrease in the time scale of rotation and increase in the relaxation time scale of the fluid tends to make the oscillations within the fluid grow and the time taken to reach steady state increase. At this juncture we would like to mention that the product of El and $\text{Re}_\Omega$ by our definition ends up to be $\text{El} \times \text{Re}_\Omega = 2\Omega\lambda_1$. This product is essentially the ratio of the relaxation time scale of the fluid to the time scale of rotation. When these two time scales are equal, we end up the product having the value 2. When the product has a value less than 2, the rotational time scale and its associated effects dominate, whereas, when the product has a value greater than 2, the fluid relaxation time scale plays the dominating role. In our observation, it is clear that for

$El \times Re_\Omega \ll 2$, the fluid tends to behave like a Newtonian fluid in rotation. The discussion for such flows are well known and can be found in literature.[15,36] When the product $El \times Re_\Omega \gg 2$, the effect of rotation is diminished and the fluid rheology tends to play the more signification role. Such flows have been well studied in literature.[16,17] In this study, we concentrate our discussion on cases where $El \times Re_\Omega \sim 2$, where the two time scales are equally important. These effects have been shown and discussed in Figs. 3(d)-(h) and Figs. 4(a)-(h).

In order to further our understanding of the flow profiles, we plot the centerline velocities $(at\ z = 1)$ in both the $x$ and $y$ directions, as depicted in Figs. 5(a) and (b) respectively. It is observed that as El increases, for both $u$ and $v$ velocities, the time up to which the centerline velocity does not feel the effect of electroosmotic forcing increases. The inset figures clearly depict that for Newtonian fluid $(i.e.\ El = 0)$ the centreline velocities develop almost instantaneously, while for the Maxwell fluid $(i.e.\ El = 1\ and\ 2)$ we observe that upto non-dimensional time $t \sim 1$ both the centerline velocities remain close to zero. This may be attributed to the increase in time scale of relaxation of the fluid with increase in El. This was also seen in Figs. 2(a) and (b), where at certain temporal instants, the velocity magnitude at the centerline became larger than the final steady state value. We also observe that there is a strong oscillatory nature of the centerline velocity with increase in El. It is important to note that for a Maxwell fluid, the diffusive nature of the fluid is not strongly prominent at smaller times, due to which we see oscillations which are not smooth in Figs. 5(a) and (b). Also the rate at which these oscillations decay is lower for higher value of El.

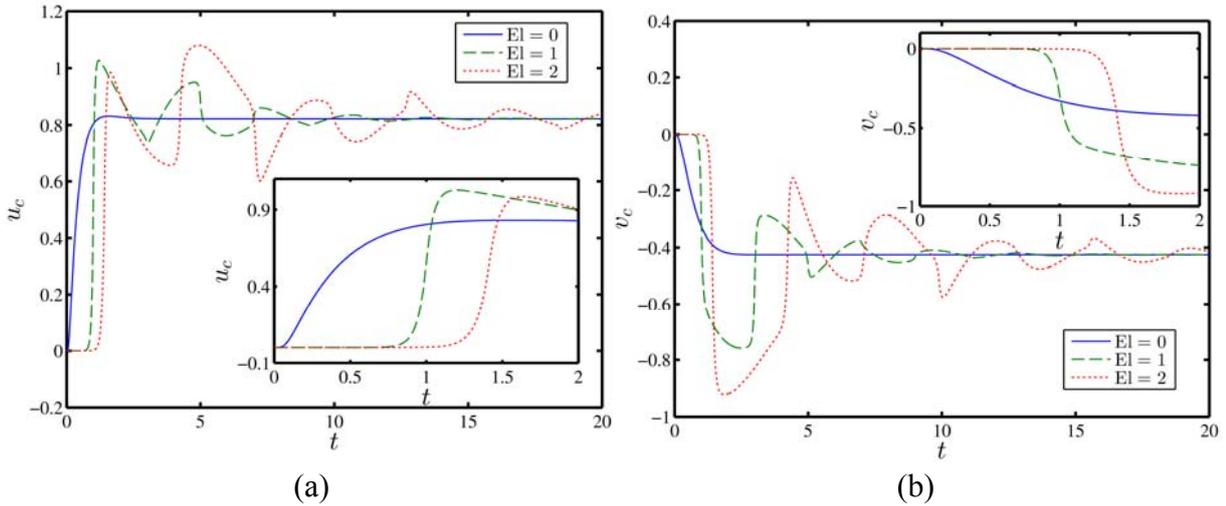

(a)            (b)

FIG. 5. Temporal evolution of (a) x component of centerline $(at\ z = 1)$ velocity and (b) y component of centerline velocity for $\kappa = 20$, $Re_\Omega = 1$ and $El = 0, 1\ and\ 2$. The insets represent the small time dynamics of the centerline velocities.

The commonly measured quantity in any flow is the flow rate, which has been defined in Eq. (51) for flows in both $x$ direction and $y$ direction respectively defined as $Q_x$ and $Q_y$. The direction of flow is given by the angle $\theta_f$ which has been defined in Eq. (51). Therefore, the temporal evolution of flow rates and direction has been depicted in Figs. 6(a)-(c). The effect of the various phases of oscillation at different locations within the fluid is captured through the flow rates, which gives a compound effect of the entire flow. For example, in Fig. 5(a) we observe that the centerline velocity magnitude is higher than steady state value at some instants of time between $t = 0$ and $t = 5$. However, in Fig 6(a), we see that the flow rate in the $x$ direction is lower than steady state value between $t = 0$ and $t = 5$. It is observed that the oscillation magnitude increases for both flow rates and direction with increase in El. Also, the time of transience also increases with increase in El. It is also observed that the oscillations in the flow rate in $y$ direction is much larger than for the flow rate in $x$ direction as depicted in Figs. 6(a) and (b). This may be attributed to the stabilizing effect that the electroosmotic body force plays on the fluid in the $x$ direction. As another artifact of the dominance of oscillation in $y$ direction, we find that the flow angle is also a strong function of the $y$ direction flow rate. One may be see a strong similarity between the plots depicted in Figs. 6(b) and (c).

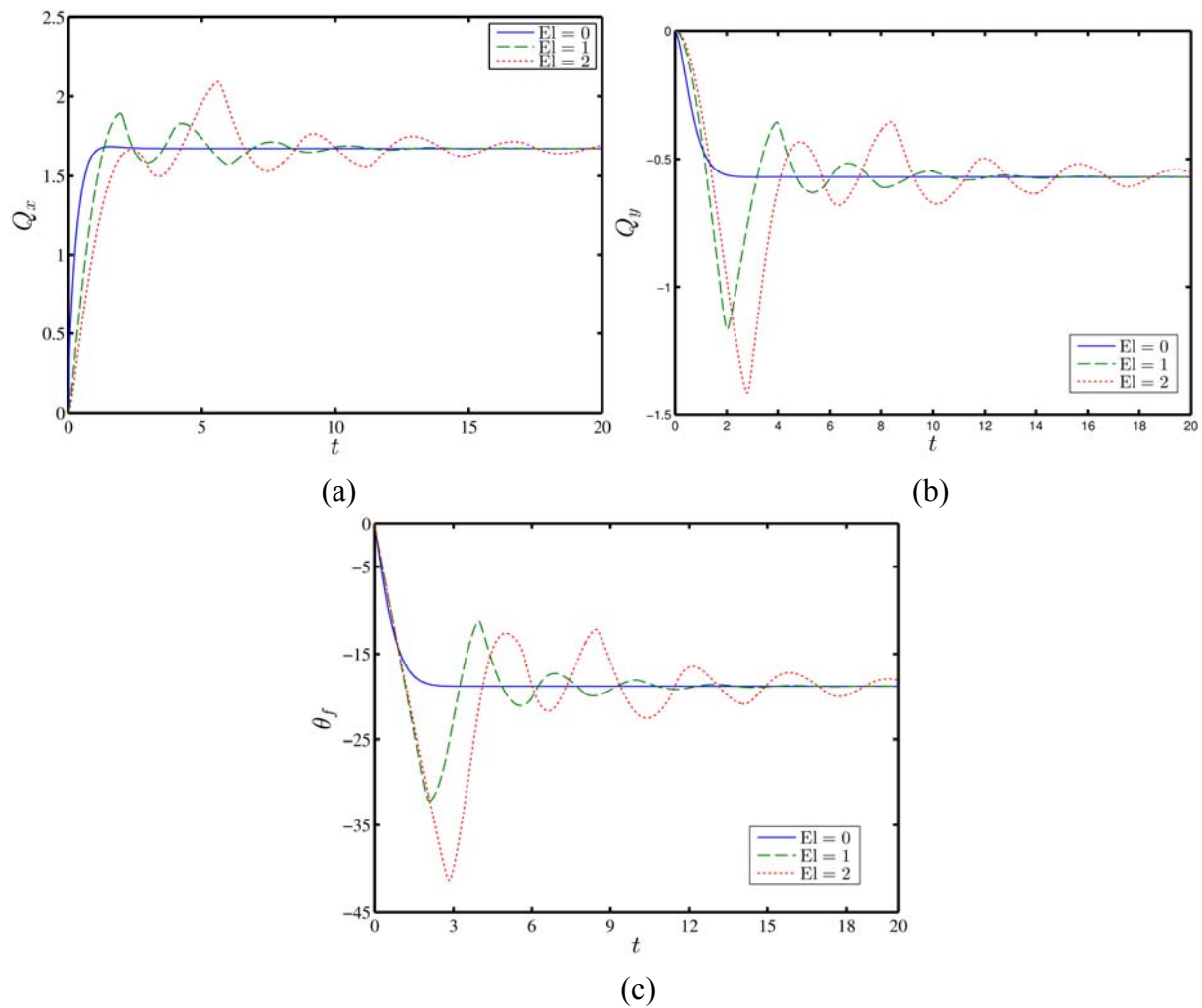

(a)  (b)  (c)

FIG. 6. Temporal evolution of (a) $x$ direction flow rate, (b) $y$ direction flow rate and (c) flow direction (angle) for $\kappa = 20$, $\text{Re}_\Omega = 1$ and $\text{El} = 0, 1,$ and $2$.

## V. CONCLUSIONS

Investigation has been conducted on the transient flow of Maxwell fluid in a rotating microfluidic channel under electroosmotic pumping by both analytical and numerical methods. We observe very good match between the analytical and numerical solutions. It has been shown that rheological effects play a significant role during transience mainly due to the interplay of time scales of relaxation of the fluid and the imposed rotational time scale. We have shown that the Coriolis force plays a significant role in altering the dynamics of electroosmotic flow of Maxwell fluid. The parameters such as rotational Reynolds number and elasticity number have been found to have a major influence on the transient flow dynamics. In general it has been found that increasing either the rotational Reynolds number or the elasticity number increases the oscillations within the fluid and also increases the time of transience. The oscillations within the fluid could be used to enhance mixing within the fluid channel for such fluids. We strongly believe that this analysis could be of potential interest in the design of bio-microfluidic devices/systems, where the associated fluids have complex rheological behaviour.

## ACKNOWLEDGEMENTS

PK and SC are indebted to the Sponsored Research Industrial Consultancy (SRIC), Indian Institute of Technology Kharagpur, for the financial support under the Project, "Center of Excellence for Research and Training in Microfluidics (CEM)" for this research.